\documentclass{elsart}
\usepackage[dvips]{graphicx}
\usepackage{amsmath}

\begin{document}
\title{Generalized thermodynamic uncertainty relations}
\author{G. Wilk$^{a}$, Z.W\l odarczyk$^{b}$}
\address{$^{a}$The Andrzej So\l tan Institute for Nuclear Studies,
                Ho\.za 69; 00-689 Warsaw, Poland\\ e-mail: wilk@fuw.edu.pl\\
         $^{b}$Institute of Physics, Jan Kochanowski University,
               \'Swi\c{e}tokrzyska 15; 25-406 Kielce, Poland;\\ e-mail:
               zbigniew.wlodarczyk@ujk.kielce.pl\\
                  }

 {\today}

{\scriptsize Abstract: We analyze an ensemble in which energy
($E$), temperature ($T$) and multiplicity ($N$) can all fluctuate
and with the help of nonextensive statistics we propose a relation
connecting all fluctuating variables. It generalizes Lindhard's
thermodynamic uncertainty relations known in literature.


}

\section{Introduction}
\label{sec:I}

A long time ago, it was suggested that the thermodynamical
quantities, temperature $T$ and energy $U$, could be regarded as
being complementary in the same way as are position and momentum
in quantum mechanics \cite{BH}. The reason is that the only way to
attribute a definite temperature to a physical system is by
bringing it into thermal contact and equilibrium with another very
large system acting as a heat bath. In this case, however, the
system will freely exchange energy with the heat bath, and one
loses the possibility of controlling its energy. On the other
hand, in order to make sure that the system has a definite energy,
one should isolate it from its environment. But then there is no
way to determine its temperature. Dimensional analysis already
leads to the conjecture that this relation would take the form $
\Delta U\, \Delta \beta \ge k $,  where $\beta = 1/T$ and $k$ is
Boltzmann's constant. The isolation ($U$ definite) and contact
with a heat bath ($T$ definite) then represent two extremal cases
of such complementarity. This idea has so far not received much
recognition in the literature because its validity and foundations
are under discussion (see \cite{UL} for review and \cite{L} for
comments). The main point is the exact meaning of the increments
$\Delta$. Indeed, all versions of this uncertainty relation
proposed so far employ different theoretical frameworks and give
different interpretations of the uncertainty $\Delta \beta$, in
most cases by using concepts from theories of statistical
inference.

In this paper we shall treat these increments as a measure of
fluctuation of the corresponding physical quantities and analyze
an ensemble in which energy ($U$), temperature ($T$) and
multiplicity ($N$), can all fluctuate. In this way we generalize
the relation between fluctuations of $U$ and $T$ derived in
thermodynamics \cite{JL} expressed by their relative variances
$Var(x)/\langle x\rangle^2 = \omega^2_x$ as:
\begin{equation}
\omega_U^2\, +\, \omega^2_T\, =\, \frac{1}{\langle N\rangle}.
\label{eq:JL}
\end{equation}
Eq. (\ref{eq:JL}) is an attempt at describing a small system
remaining in thermal contact with a heat bath of varying size. It
represents the kind of uncertainty relation mentioned before,
namely that the standard deviation of one variable can be made
small only at the expense of increasing the corresponding standard
deviation of the conjugate variable \cite{UL,L}. This relation is
supposed to be valid all the way from the canonical ensemble, for
which $Var(T) = 0$ and $Var(U) = 1/\langle N\rangle$, up to the
microcanonical ensemble for which $Var(T) = 1/\langle N\rangle$
and $Var(U) = 0$. It expresses the complementarity between the
temperature and energy and the canonical and microcanonical
description of the system. The generalization proposed here
extends relation (\ref{eq:JL}), also including in it the possible
fluctuations of multiplicity $N$ and its connections with
fluctuations of temperature; this is done using nonextensive
Tsallis statistics \cite{T} which incorporates such fluctuations
in a natural way \cite{WW,BJ}. We shall demonstrate that such an
approach results in a characteristic relation connecting all
fluctuating variables.

\section{Fluctuations  in statistical mechanics}
\label{sec:II}

\subsection{Single-variable fluctuations}

Suppose that in a process one has $N$ independently produced
secondaries with energies $\left\{ E_{i=1,\dots,N}\right\}$, each
distributed according to Boltzmann distribution characterized by a
temperature $T$ \footnote{In the relativistic regime, where masses
are negligible in comparison with momenta, $p << m$, and for the
cylindrical phase space in which $p_L
>> p_T$, we have $E \cong p_L$. Therefore, when replacing
discrete values $E_i$ in the energy distribution function, Eq.
(\ref{eq:BG}), by a continuous variable $E$ (and replacing
summations by integrals), we can put, for simplicity, the density
of states $\rho(E) = 1$. In this case the energy distribution
function (\ref{eq:BG}) is normalized to unity. This simplification
leads to simple interpretation of temperature parameter, $T =
\langle E\rangle$, and specific heat, $c=1$.},
\begin{equation}
g_i\left(E_i\right) = \beta\cdot \exp\left( - \beta\cdot
E_i\right)\quad {\rm where}\quad \beta^{-1} = T. \label{eq:BG}
\end{equation}
The corresponding joint probability distribution in this case is
given by
\begin{equation}
g\left( \left\{ E_{i=1,\dots,N}\right\} \right) = \beta^N\,
\exp\left( - \beta\, \sum_{i=1}^N E_i \right) = \beta^N\, \exp( -
\beta\,U), \qquad U = \sum_{i=1}^N E_i .\label{eq:sumBG}
\end{equation}

When two of three variables $(U,T,N)$ are fixed, one has three
possible situations:
\begin{itemize}
\item  $T = 1/\beta$ and $N$ are fixed and the energy $U$ (because
$\left\{ E_{i=1,\dots,N}\right\}$ are independent) can fluctuate.
Using characteristic functions or sequentially performing
integration of the joint distribution (\ref{eq:sumBG}) and
noticing that,
\begin{equation}
g_{T,N}(U) = g_{T,N-1}(U)\cdot \frac{U}{(N-1)}, \label{eq:seq}
\end{equation}
one obtains that the energy $U$ fluctuates according to gamma
distribution:
\begin{eqnarray}
g_{T,N}(U) &=& \frac{\beta}{\Gamma(N)} (\beta U)^{N-1} \exp(-\beta
U
),\label{TN1}\\
&&\frac{Var(U)}{\langle U\rangle^2} = \frac{1}{N}, \label{eq:TN}
\end{eqnarray}
where $\langle U\rangle = T N$ and $Var(U) = \langle U^2 \rangle -
\langle U\rangle^2 = T^2 N(N+1) - T^2N^2 = T^2 N$.

\item $T = 1/\beta$ and $U$ are fixed and the multiplicity $N$
fluctuates. In this case one first writes a cumulative
distribution function for the probability density function given
by Eq. (\ref{eq:TN}),
\begin{equation}
G_{T,N}(U) = 1\, - \, \sum_{i=1}^{N-1}\frac{1}{(i - 1)!} ( \beta U
)^{i-1}\exp( - \beta U). \label{eq:GTN}
\end{equation}
For energies $\left\{ E_{i=1,\dots,N}\right\}$ such that
\begin{equation}
\sum^N_{i=0} E_i \leq U \leq \sum^{N+1}_{i=0}E_i,
\label{eq:condition}
\end{equation}
the corresponding multiplicity distribution (notice that $U/T =
\langle N\rangle $) is
\begin{equation}
g_{T,U}(N) = G_{T,N+1}(U) - G_{T,N}(U) = \frac{\langle
N\rangle^N}{N!}\, \exp( - \langle N\rangle), \label{eq:TU}
\end{equation}
i.e., it has form of the Poisson distribution with
\begin{equation}
\frac{Var(N)}{\langle N\rangle^2} = \frac{1}{\langle N\rangle}.
\label{eq:VarP}
\end{equation}

\item $N$ and total energy $U$ are fixed and temperature $T$
fluctuates. Inverting distribution $g_{T,N}(U)$, Eq.
(\ref{eq:TN}), one gets\footnote{~~Usually the temperature
fluctuations in a system are related to its heat capacity under
constant volume \cite{LDL,LS}, $C_V$, by $\omega^2_{\beta} =
1/C_V$ and $C_V = c\langle N\rangle$ (in what follows we shall put
specific heat $c=1$). It means therefore that $\omega_T^2 \cong
\omega^2_{\beta} = 1/\langle N\rangle$.}

\begin{eqnarray}
g_{U,N}(T) &=& \frac{\partial }{\partial \beta}\,\int^U_0 dU'\,
g_{T,N}(U') = \frac{U}{\Gamma(N)} (\beta U)^{N-1}\, \exp( - \beta
U), \label{eq:UN}\\
&& \frac{Var(T)}{\langle T\rangle^2} \cong
\frac{Var(\beta)}{\langle \beta\rangle^2} = \frac{1}{N},
\label{eq:UNvar}
\end{eqnarray}
where $\langle \beta\rangle = N/U$and $Var(\beta) = \langle
\beta^2\rangle - \langle \beta\rangle^2 = N(N + 1)/U^2 - N^2/U^2 =
N/U^2$.
\end{itemize}

Notice that all limiting distributions, Eqs. (\ref{eq:TN}),
(\ref{eq:TU}) and (\ref{eq:UN}), have the formally identical form
of a gamma distribution,
\begin{equation}
U g_{T,N}(U)\, =\, N g_{T,U}(N)\, =\, \beta g_{U,N}(\beta) =
\frac{(\beta U)^N}{\Gamma(N)}\, \exp( - \beta U), \label{eq:all}
\end{equation}
and also have identical respective relative fluctuations,
\begin{equation}
\frac{Var(U)}{\langle U\rangle^2}\, =\, \frac{Var(\beta )}{\langle
\beta\rangle^2}\, =\, \frac{Var(N)}{\langle N\rangle^2}\, =\,
\frac{1}{\langle N\rangle} .\label{eq:allfluct}
\end{equation}
For large $N$ Eq. (\ref{eq:all}) becomes a Gaussian distribution,
\begin{equation}
g(x) = \theta\frac{(\theta x)^{N-1}}{\Gamma(N)}\, \exp( - \theta
x)\, \stackrel{N\rightarrow \infty}{\longrightarrow}\,
\frac{1}{\sqrt{2\pi}\sigma}\, \exp\left( - \frac{(x -
\mu)^2}{2\sigma^2} \right) \label{eq:Gauss}
\end{equation}
which (with $\mu = N/\theta$ and $\sigma^2 = N/\theta^2$) is the
distribution usually used to describe fluctuations in statistical
physics \cite{LDL}.

\subsection{Generalized fluctuations}

Our considerations can be generalized by resorting to Tsallis
statistics \cite{T} with fluctuating $U$ and $T$. In \cite{WW,BJ}
it was shown that fluctuations of temperature in a heat bath in
the form of a gamma distribution, result in a Tsallis
distribution,
\begin{equation}
h_q(E) = \exp_q \left(-\frac{E}{T}\right) = \frac{2 - q}{T}\left[1
- (1-q)\frac{E}{T}\right]^{\frac{1}{1-q}}, \label{eq:Tsallis}
\end{equation}
with one new parameter, a nonextensivity parameter $q$ ($q \leq
2$). The Boltzman distribution, Eq. (\ref{eq:BG}), is recovered
for $q \rightarrow 1$. It turns out \cite{WW} that the
nonextensivity parameter $q$ in Eq. (\ref{eq:Tsallis}) is given by
these fluctuations of temperature $T$:
\begin{equation}
\frac{Var(T)}{\langle T\rangle^2} = \omega^2_T = q - 1.
\label{eq:defq}
\end{equation}
The further consequence of using Tsallis statistics is that now
the joint $N$-particle Tsallis distribution with energies $\{
E_{i=1,\dots,N}\}$,
\begin{equation}
h\left(\{ E_{i=1,\dots,N}\} \right)\! =\! C_N\left[ 1-
(1-q)\frac{\sum^N_{i=1} E_i }{T} \right]^{\frac{1}{1-q}+1-N},
\label{eq:NTsallis}
\end{equation}
does not factorize into single particle distributions as $g\left(
\left\{ E_{i=1,\dots,N}\right\} \right)$ in Eq. (\ref{eq:sumBG})
\cite{fluct}. As a result, the corresponding multiplicity
distribution, which in the case of Boltzman-Gibbs statistics has a
Poissonian form, cf. Eq. (\ref{eq:TU}), now takes a Negative
Binomial (NB) form \cite{fluct},
\begin{equation}
P(N)\, =\, \frac{\Gamma(N+k)}{\Gamma(N+1)\Gamma(k)}\frac{\left(
\frac{\langle N\rangle}{k}\right)^N}{\left( 1 + \frac{\langle
N\rangle}{k}\right)^{(N+k)}}, \label{eq:NBD}
\end{equation}
where parameter $k$ is given by the parameter $q$ from Eq.
(\ref{eq:Tsallis}),
\begin{equation}
k = \frac{1}{ q - 1} \label{eq:kq}
\end{equation}
(in what follows we only consider the case of $q \ge 1$). On the
other hand, from the definition of NB distribution (\ref{eq:NBD}),
\begin{equation}
\frac{1}{k} = \frac{Var(N)}{\langle N\rangle^2} - \frac{1}{\langle
N\rangle} = \omega_N^2 - \frac{1}{\langle N\rangle}. \label{eq:qN}
\end{equation}
It means that fluctuations of $N$ and $T$ are not independent, but
related in the following way:
\begin{equation}
\omega_N^2 - \frac{1}{\langle N\rangle} = \omega_T^2.
\label{eq:relNT}
\end{equation}

However, NB multiplicity distribution can be obtained also as a
result of fluctuations (in the form of a gamma function) of the
mean multiplicity, $\bar{N} = \langle N\rangle$, in the Poisson
distribution. That is because in this case \cite{CCS} (cf., also
\cite{NUWW,VP}):
\begin{equation}
P(N)\, =\, \int_0^{\infty} d\bar{N}
\frac{e^{-\bar{N}}\bar{N}^N}{N!}\cdot
         \frac{\gamma^k \bar{N}^{k-1} e^{-\gamma \bar{N}}}{\Gamma (k)} =
   \frac{\Gamma(k+N)}{\Gamma (1+N) \Gamma (k)}\cdot
   \frac{\gamma^k}{(\gamma +1)^{k+N}} \label{eq:PNBD}
\end{equation}
that, for $\gamma = \frac{k}{\langle \bar{N}\rangle}$, coincides
with Eq. (\ref{eq:NBD}). Therefore, in addition to Eq.
(\ref{eq:qN}) one also has that
\begin{equation}
\frac{1}{k} = \frac{Var(\bar{N})}{\langle \bar{N}\rangle^2}.
\label{eq:anotherNB}
\end{equation}
For $\bar{N} = const$ we have a Poisson distribution ($ k =
\infty$). Fluctuating $1/T$ according to gamma distribution and
keeping $U = const$ results in $Var(\bar{N})/\langle
\bar{N}\rangle^2 = \omega^2_T$ and we recover Eq.
(\ref{eq:relNT}). Analogously, fluctuating $U$ while keeping
$T=const$ gives us $Var(\bar{N})/\langle \bar{N}\rangle^2 =
\omega^2_U$.

Fluctuating both $U$ and $T$ (and taking into account  that
$\bar{N} = U/T$) one has that
\begin{equation}
\frac{Var(\bar{N})}{\langle \bar{N}\rangle^2} =
Var\left(\frac{U}{T}\right)\cdot \left(\frac{\langle
T\rangle}{\langle U\rangle}\right)^2 \cong \left[
\frac{Var(U)}{\langle U \rangle^2} + \frac{Var(T)}{\langle
T\rangle^2} - 2\frac{Cov(U,T)}{\langle U\rangle\langle T\rangle}
\right]\label{eq:covN}
\end{equation}
or, in terms of the scaled variances introduced before,
\begin{equation}
\omega^2_{\bar{N}} \cong \omega^2_U\, +\, \omega^2_T\, -\, 2\rho
\omega_U\omega_T, \label{eq:correl}
\end{equation}
where $\rho = \rho (U,T)$ is the correlation coefficient ($\rho
\in [-1,1]$)\footnote{~~~A similar relation connecting variables
$V$, $P$ and $T$, $\omega^2_P = \omega^2_V + \omega^2_T$, is known
for almost a century \cite{Q1}.}.

Comparing Eqs. (\ref{eq:qN}) and (\ref{eq:anotherNB}) and
accounting for (\ref{eq:correl}) one gets the following general
relation between all fluctuating variables:
\begin{equation}
\Big| \omega^2_N - \frac{1}{\langle N\rangle}\Big| = \omega^2_U +
\omega^2_T - 2\rho \omega_U \omega_T. \label{eq:corq}
\end{equation}
This relation, which is our main result, generalizes Linhard's
thermodynamic uncertainty relation given by Eq. (\ref{eq:JL}).

A word of explanation is in order. The use of $|\dots|$ makes our
formula (\ref{eq:corq}) general, i.e., valid for both $\omega^2_N
\ge 1/\langle N\rangle$ and for $\omega^2_N =0$ if $N=const$.
Actually, when all variables fluctuate one cannot have
fluctuations of $N$ smaller than the Poissonian. Observation of
sub-Poissonian fluctuations, which would correspond to the case $q
< 1$, always signal the presence of some additional constraints
(like conservation of some quantum numbers, for example charges,
cf., \cite{Gor}). We restrict ourselves to the case $q \ge 1$ and
do not describe the region $0 < \omega^2_N < 1/\langle N\rangle$.
We could, therefore, alternatively write Eq. (\ref{eq:corq}) as
\begin{equation}
\omega^2_N - \frac{1}{\langle N\rangle} \left(1 - 2 \delta\left(
\omega^2_N\right) \right) = \omega^2_U + \omega^2_T - 2\rho
\omega_U \omega_T, \label{eq:footnote}
\end{equation}
where $\delta(x)$ is Dirac delta. However, in what follows, we
shall use Eq. (\ref{eq:corq}).

It is straightforward to see that when two of three variables are
fixed results obtained using relation (\ref{eq:corq}) coincide
with those obtained using Boltzman statistics (cf. Eq.
(\ref{eq:allfluct})). When only one variable is kept constant we
have:
\begin{itemize}
\item for $U = const$ one gets Eq. (\ref{eq:relNT})  obtained
using Tsallis statistics;

\item for $T = const$ one gets
\begin{equation}
\omega^2_N - 1/\langle N\rangle = \omega^2_U, \label{eq:relNU}
\end{equation}
 i.e., fluctuations of energy result in fluctuations in
 multiplicity identical to those induced by fluctuations in
 temperature, cf. Eq. (\ref{eq:relNT});

\item for $N = const$ one gets $ 1/N = \omega^2_U + \omega^2_T - 2
\rho \omega_U \omega_T$, where for $\rho = 0$ one obtains Eq.
(\ref{eq:JL}) proposed in \cite{L}; because $-1 \leq \rho \leq 1$
the relative fluctuations are constrained by the relation $\left(
\omega_U - \omega_T\right)^2 \leq 1/N \leq \left( \omega_U +
\omega_T\right)^2$.
\end{itemize}

\begin{figure}[h]
  \begin{center}
   \includegraphics[width=11.0cm]{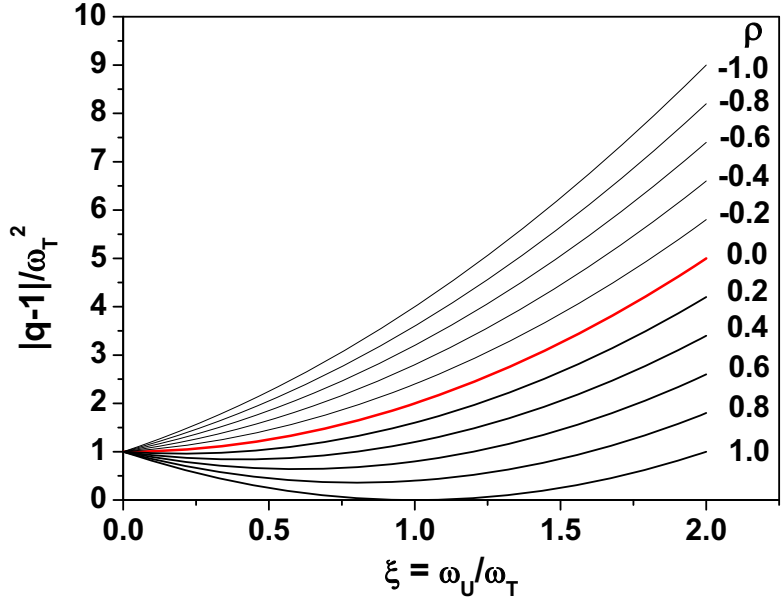}
   \caption{(Color online) Dependence of the ratio $|q - 1|/\omega^2_T$
             on the relative fluctuations, $\xi$, for different
             values of correlation parameter $\rho$.}
   \label{Figure1}
  \end{center}
\end{figure}

For the case when all variables are free to fluctuate we have Eq.
(\ref{eq:corq}) which can be rewritten as:
\begin{equation}
\Big| \omega^2_N - \frac{1}{\langle N\rangle}\Big| = \left(
\omega_U - \omega_T\right)^2 + 2 \omega_U \omega_T (1 - \rho).
\label{eq:newcorq}
\end{equation}
A Poissonian distribution of multiplicity (i.e., $\omega^2_N = 1/
\langle N\rangle$) is possible only for $\rho = 1$ when $\omega_U
= \omega_T$. For $\rho = 0$ one has $| \omega^2_N - \langle
N\rangle^{-1}| = \omega^2_U + \omega^2_T$. Using Eqs.
(\ref{eq:newcorq}) and (\ref{eq:qN}) one can express the
nonextensivity parameter $q$ by the respective fluctuations and
correlations
 and write
\begin{equation}
| q - 1 | = \left( \omega_U - \omega_T\right)^2 + 2 \omega_U
\omega_T ( 1 - \rho ) = \omega^2_T \left[(1 - \xi)^2 + 2 \xi (1 -
\rho)\right],\label{eq:qNg}
\end{equation}
where $\quad \xi = \omega_U/\omega_T$. Dependence of $|q -
1|/\omega^2_T$ on the correlation coefficient, $\rho$, and
relative fluctuations, $\xi$, is shown in Fig. \ref{Figure1}.
When $U = const$ (i.e., $\xi = 0$) we have the same situation we
encountered for a fluctuating temperature, namely that $| q - 1
|/\omega_T^2 = 1$. When we allow energy to fluctuate and add these
fluctuations, $ |q - 1|/\omega^2_T$ changes as shown in Fig.
\ref{Figure1}. Notice that condition $q \le2$, implies that
$\omega_U + \omega_T \leq 1$, for $\rho \neq 0$, and that
$\omega^2_U + \omega^2_T \leq 1$ for $\rho = 0$ \footnote{~~~~~
When energy $U$ fluctuates, the pairs of variables, $(U,N)$ and
$(U,T)$, cannot be independent simultaneously because $Var(U) =
\langle T\rangle Cov(U,N) + \langle N\rangle Cov(U,T)$ . This
relation arises from a comparison of Eq. (\ref{eq:covN})) with the
analogous formula evaluated for variable $T = U/N$.}.

\section{An application}
\label{sec:III}

As an example we compare fluctuations extracted from the
distribution of different observables in a high energy
multiparticle production process. It should be remembered that Eq.
(\ref{eq:corq}) connects fluctuations of different observables,
but defined in the same fragment of allowed phase space, whereas
available data usually refer to different parts of this phase
space. Therefore, corresponding $q$ parameters are usually
difficult to compare. For example, $q = q_L$ obtained from
rapidity ($y$) distributions, $dN/dy$, defined in so-called
longitudinal phase space, are comparable with q evaluated from the
multiplicity distributions, $P(N)$, which are defined in the full
phase space \cite{NUWW,WWW}. On the other hand, transverse
momentum ($p_T$) distributions, $dN/dp_T$, defined in the
so-called transverse space, are described by much smaller values
of $q = q_T$. A first attempt to explore the relation
(\ref{eq:JL}) in high energy multiparticle production processes
was presented in \cite{GWZW}.

\begin{figure}[h]
  \begin{center}
   \includegraphics[width=14.0cm]{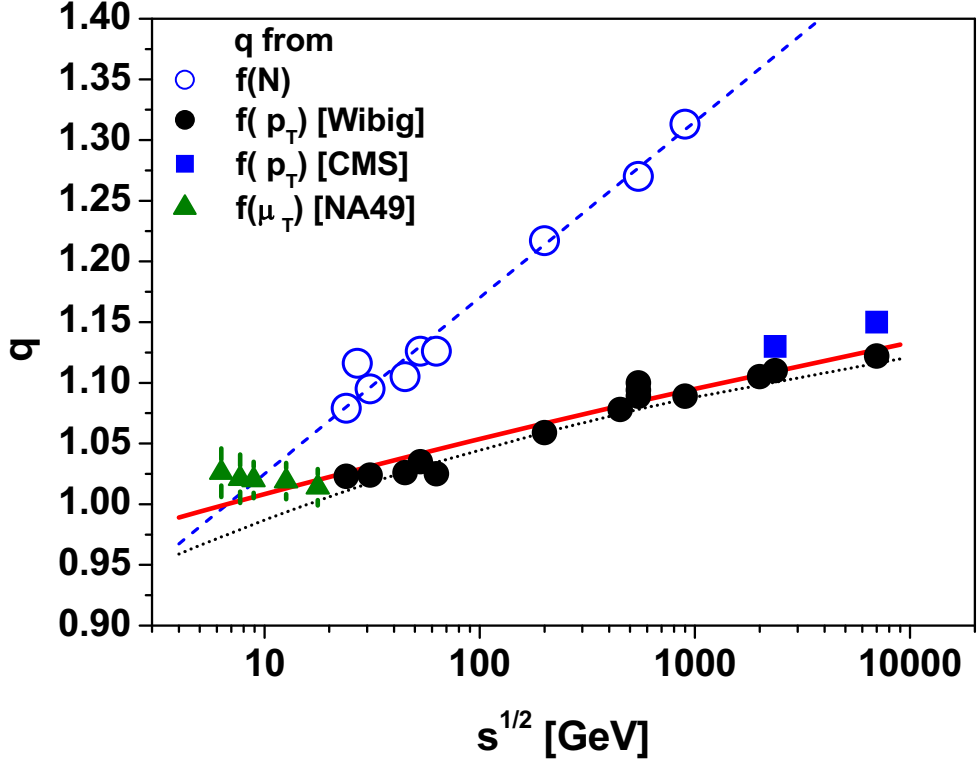}
   \caption{(Color online) An example of energy dependencies of
             the nonextensivity parameters $q$ obtained from
             different observables. Open symbols: $q$
             obtained from multiplicity distributions $P(N)$
             \cite{PN} (fitted by $q = 1 + 1/k$ with $1/k= -0.04
             + 0.029\ln s$, cf. \cite{PN}). Closed symbols:
             $q = q_T$ obtained from a different analysis of
             transverse momenta distributions, $f \left( p_T \right)$.
             Data points are from, respectively, \cite{Wibig}
             for [Wibig], \cite{CMS} for [CMS] and from \cite{NA49}
             (data on $\mu_T = \sqrt{m_{\pi}^2 + p_T^2}$)
             for [NA49]). The dotted line represents a fit
             obtained in \cite{Wibig} ($q_T = 1.25 -
             0.33s^{-0.054}$) and the full line comes from
             our Eq. (\ref{eq:qqTn}).}
   \label{Figure2}
  \end{center}
\end{figure}
So far there are no data which would necessitate the use of
nonzero correlations. In the case of uncorrelated fluctuations
($\rho = 0$ ), one gets from Eq. (\ref{eq:corq}), using
(\ref{eq:qN}), that
\begin{equation}
\frac{1}{k} = q - 1 = \omega^2_U + \omega^2_T\, .
\label{eq:OMUOMT}
\end{equation}

In Fig. \ref{Figure2} we plot the energy dependencies of the
nonextensivity parameter $q$ obtained from different sources: from
the multiplicity distributions, $f(N) = P(N)$, i.e., from the full
phase space \cite{PN} and from a different analysis of transverse
momenta distributions, $f\left( p_T\right)$ (i.e., from the
transverse phase space \cite{Wibig,CMS,NA49}). The characteristic
feature seen there is that, whereas the former show substantial
energy dependence (and essentially follow results for $q = q_L$
obtained in \cite{NUWW} from the analysis of $dN/dy$), the latter
$q = q_T$ are only weakly dependent on the interaction energy
(notice also that $q_T(s)$ from different sources plotted in Fig.
\ref{Figure2} are roughly the same). To somehow compare $q$ and
$q_T$, one needs some additional input. Assume that, being only
weakly energy dependent, $q_T$ are also roughly independent of the
energy fluctuations. It is now natural to expect that transverse
characteristics are mainly governed by the fluctuations of
temperature, i.e., that we can write
\begin{equation}
q_T - 1 = \frac{2}{3} \omega^2_T \label{eq:qT}
\end{equation}
(it is assumed here that fluctuations of temperature contribute
equally to each of the components of momenta, hence the factor
$2/3$). Following this line of thought, i.e., assuming
additionally that fluctuations of energy $U$ are entirely given by
its thermal part, one can write that in this case $\omega^2_U =
\omega^2_T$ and that both parameters, $q$ and $q_T$, are connected
by the following relation:
\begin{equation}
q - 1 = 3 \left( q_T - 1\right)\qquad{\rm or}\qquad q_T =
\frac{1}{3}(2 + q). \label{eq:qqT}
\end{equation}

A few explanatory remarks are in order. Namely, in
\cite{Wibig,CMS} the $p_T$ distributions were fitted using Tsallis
distributions using the power $1/(1 - q_T)$. However, one should
keep in mind that in $p_T$ distributions one has the prefactor
$p_T/T^2$, not $1/T$, present in the usual exponential
distributions. This fact results in a slightly different power,
$q_T/(1 - q_T)$ \cite{fluct} (this is the situation similar to the
change from the so-called superstatistics A, in which only
expression under the exponent is subjected to fluctuations, to
superstatistics B, in which one fluctuates also the prefactor
\cite{fluct}, cf. also \cite{SS}). As a result the relation
between $q_T$ and $q$ is slightly modified and Eq. (\ref{eq:qqT})
now becomes
\begin{equation}
q_T = \frac{1 + 2q}{2 + q}. \label{eq:qqTn}
\end{equation}
Using values of $q = 1 + 1/k = 0.896 + 0.029\ln s$ obtained from
$P(N)$ \cite{PN}, the evaluated $q_T$ is shown in Fig.
\ref{Figure2} (solid line) and compared to $q_T$ extracted from
transverse momenta distributions $f(p_T)$ \cite{Wibig,CMS,NA49}.
This, in turn, should be compared with the dotted line
representing the fit in \cite{Wibig} resulting in $q_T = 1.25 -
0.33s^{-0.054}$. Notice the good agreement of Eq. (\ref{eq:qqTn})
with data which, in our opinion, justifies the statement that Eq.
(\ref{eq:qqTn}) represents a kind of rule connecting fluctuations
in different parts of phase space (modulo additional assumptions).

\section{Summary}
\label{sec:IV}

Using nonextensive statistics applied to ensembles in which the
energy ($E$), temperature ($T$) and multiplicity ($N$) fluctuate,
we have derived a specific relation connecting all fluctuating
variables, Eq. (\ref{eq:corq}), which generalizes Linhard's
thermodynamic uncertainty relation given by Eq.
(\ref{eq:JL})\footnote{~~~~~~~Actually, the nonextensive approach
used here is still subject to a debate about whether it is
consistent with the equilibrium thermodynamics \cite{debate}. In
this respect we would like to say that recently it was
demonstrated on general grounds \cite{M} that fluctuation
phenomena can be incorporated into traditional presentation of
thermodynamic and that the Tsallis distribution (\ref{eq:Tsallis})
belongs to the class of general admissible distributions which
satisfy thermodynamical consistency conditions and are a natural
extension of the usual Boltzman-Gibbs canonical distribution
(\ref{eq:BG}). Still other justification of a nonextensivity
approach can be found in \cite{BUS}.}. This is illustrated using
example taken from the multiparticle production processes. A
possibility of connecting fluctuations appearing in different
parts of phase space is indicated, cf. Eq. (\ref{eq:qqTn}).

\section*{Acknowledgements}
Partial support (GW) of the Ministry of Science and Higher
Education under contract DPN/N97/CERN/2009 is acknowledged.

\end{document}